\documentstyle[12pt]{article}
\begin{document}
\baselineskip=24ptplus.5ptminus.2pt
\setcounter{page}{1}
\vspace*{0.5 in}
\large

\begin{center}
Quark-Meson Coupling Model for a Nucleon 
\end{center}

\normalsize

\begin{center}
S. W. Hong$^{a }$, B. K. Jennings$^{b}$
\par
$^{a}$
Department of Physics and Institute of Basic Science, Sungkyunkwan University,
Suwon 440-746, Korea  \\
$^{b}$
TRIUMF, 4004 Wesbrook Mall, Vancouver, British Columbia, Canada V6T 2A3 \\
\end{center}
\vspace{2ex}

\begin{center}
ABSTRACT
\end{center}

The quark-meson coupling model for a nucleon is considered. The
model describes a nucleon as an MIT bag, in which quarks are
coupled to scalar and vector mesons. A set of coupled
equations for the quark and the meson fields are obtained and are solved
in a self-consistent way. It is shown that the mass of a nucleon
as a dressed MIT bag interacting with $\sigma$- and $\omega$-meson fields
significantly differs from the mass of a free MIT bag. A few
sets of model parameters are obtained  so that the mass of a
dressed MIT bag becomes the nucleon mass.
The results of our calculations imply that the self-energy of the
bag in the quark-meson coupling model is significant and needs to
be considered in doing the nuclear matter calculations.

\par
\vspace{2ex}
PACS number: 24.85.+p, 21.65.+f, 12.39.Ba

\section{Introduction}
Quantum hadrodynamics  \cite{SW} proved to be a very
successful model in describing the bulk properties of nuclear
matter as well as the properties of finite nuclei. The model is
rather simple with only a few parameters, and yet it has been
successfully applied to a vast number of nuclear matter and
nuclear structure problems. On the
other hand, this model has a fundamental shortcoming that the
nucleon with a composite structure is treated as a point particle.
Several years ago a model
to remedy this problem was proposed by Guichon \cite{Guichon}.
He proposed a quark-meson coupling (QMC) model and
investigated the direct quark degrees of freedom in 
producing the nuclear saturation mechanism in nuclear matter.
The model describes nuclear matter as
non-overlapping MIT bags \cite{MIT,Bagmod} interacting through the
self-consistent exchange of scalar and vector mesons in the
mean-field approximation. In other words, the nucleon in nuclear matter
is assumed
to be a static spherical MIT bag in which quarks interact with the
scalar ($\sigma$)- and the vector ($\omega$)-meson fields. The model was
refined later to include the nucleon Fermi motion and the
center-of-mass corrections to the bag energy \cite{Fleck} and was
applied to a number of problems [6 $-$ 11].

In the MIT bag model \cite{MIT,Bagmod} the bag constant $B$ and a
phenomenological parameter $Z$ are fixed such that the nucleon
mass of 939 MeV is reproduced for some bag radius $R$. The bag
constant $B$ produces the pressure to make a bubble in the QCD
vacuum. $Z$ is to account for various corrections including the
zero-point motion. In the QMC model  
the quark-$\sigma$ coupling constant
($g^q_\sigma$) and the quark-$\omega$ coupling constant
($g^q_\omega$) are introduced. Usually, $B$ and $Z$ are fixed first
so that the mass of the {\it free} MIT bag becomes equivalent to the 
the nucleon mass of 939 MeV for a certain bag radius.
$B^{free}$ and $Z^{free}$ thus obtained are used 
in the QMC model calculations for nuclear matter problems. 
$g_{\sigma}^{q}$ and $g_{\omega}^{q}$ are then determined 
so as to reproduce the binding energy per nucleon (B.E. = $-$16 MeV) 
at the saturation density ($\rho_N^0 =$ 0.17 fm$^{-3}$).

However, due to the interaction between the quarks and the
$\sigma$- and the $\omega$-mesons it is expected
that the mass of a {\it single} dressed MIT bag in free space,
when calculated with $B^{free}$ and $Z^{free}$ described above,
would no longer be the nucleon mass of 939 MeV. 
Such a deviation of the mass of the dressed bag in free space 
has been neglected in previous QMC calculations in
choosing the model parameters.     
If the deviation in mass is significant, it is necessary to modify
the parameters to get the correct
nucleon mass before implementing them in the nuclear
matter calculations. It is this change in the mass of a  
dressed MIT bag that we address ourselves to in this paper.
In Section 2 we describe the QMC model for a nucleon.
A brief review of the QMC model for nuclear matter is given in Section 3.
Section 4 contains the results of the QMC model calculations for 
a nucleon and nuclear matter.
The paper is summarized in Section 5.

\section{Quark-meson coupling model for a nucleon}

The Lagrangian density for the MIT bag in which quark fields $\psi$
are coupled to the $\sigma$- and $\omega$- fields may be written as
\begin{eqnarray}
{\cal L } &=& \left[ \frac{i}{2} \left( \bar{\psi} \gamma^\mu \partial _\mu \psi
-(\partial _\mu \bar{\psi}) \gamma^\mu \psi \right)
-\bar{\psi} \left( g^q_\omega \gamma_\mu \omega^\mu + 
( m_q - g_\sigma^q \sigma ) \right)
\psi -B \right] \theta_v \nonumber \\
& & -\frac{1}{2}\bar{\psi}\psi \Delta_s - \frac{1}{4}F_{\mu\nu}F^{\mu\nu}
+\frac{1}{2}m^2_\omega \omega_\mu \omega^\mu
+\frac{1}{2} \left[ (\partial_\mu \sigma) (\partial^\mu \sigma) 
-m_{\sigma}^2 \sigma^2 \right] ,
\label{Lag}
\end{eqnarray}
where $F_{\mu\nu} = \partial_\mu \omega_\nu - \partial_\nu \omega_\mu $,
$\theta_v$ is the step function for confining the quarks inside the bag,
and $\Delta_s$ is the $\delta$-function at the bag surface.
We will neglect the isospin breaking and take $m_q = (m_u + m_d )/2$ hereafter.
For numerical calculations $m_q$ will be taken to be zero.
The mass of $\sigma$ ($\omega$) is taken as 550 (783) MeV.
From this Lagrangian density the equations of motion
for quark field $\psi$, sigma field $\sigma$,
and omega field $\omega_\mu$ follow.
For $\psi$ we have 
\begin{equation}
\left[ \gamma^\mu (i \partial_\mu - g_\omega^q \omega_\mu ) - (m_q - g_\sigma^q \sigma ) \right]
\psi \theta_v = \frac{1}{2} ( 1- i \gamma^\mu n_\mu ) \psi \Delta_s.
\label{eq:psi}
\end{equation}
The left hand side of this equation gives us
the equation for the quarks inside the bag, and
the right hand side provides us with the linear boundary condition at the
bag surface ($r = R$). 
The equations for $\sigma$ and $\omega_\mu$ are, respectively,
\begin{equation}
\partial_\mu\partial^\mu \sigma + m_\sigma^2 \sigma =
 g_\sigma^q \bar{\psi}\psi  \theta_v
\label{eq:sigma}
\end{equation}
and
\begin{equation}
\partial^\nu F_{\nu\mu} + m^2_{\omega} \omega_\mu =
g_\omega^q \bar{\psi} \gamma_\mu \psi \theta_v .
\label{eq:omega}
\end{equation}

If we consider only the ground state quarks in the static spherical MIT bag
and keep only the time-component of $\omega_{\mu}$,
Eq. (\ref{eq:psi}) becomes the following
coupled linear differential equations for $g(r)$ and $f(r)$ inside the bag;
\begin{eqnarray}
\frac{d f(r)}{dr} &=& -\left[ 2\frac{f(r)}{r} + E^- (r) g(r)
\right] \label{eq:psi1} \\ 
\frac{dg(r)}{dr} &=& E^+ (r) f(r),
\label{eq:psi2}
\end{eqnarray}
where $g(r)$ and $f(r)$ are the radial parts of 
the upper and the lower components of $\psi$, $i.e.,$
\begin{equation}
\psi(t,{\bf r}) \; = \; e^{-i\epsilon_q ^0 t/R} 
\left( 
\begin{array}{c}
             g(r) \\
  - i {\bf \vec{\sigma} } \cdot {\bf \hat{r}} f(r) \\
\end{array}  
\right) \frac{\chi_q }{\sqrt{4\pi}},
\label{eq:psi-tr}
\end{equation}
where ${\bf \vec{\sigma} }$ is the Pauli spin matrix and 
$\chi_q$ is the quark spinor. 
Also,
\begin{eqnarray}
E^+ (r) &=& \frac{\epsilon_q ^0}{R} - g_{\omega}^q \omega_0 ( {\bf r} ) 
         + (m_q -g_\sigma^q \sigma ( {\bf r} ))\nonumber \\
E^- (r) &=& \frac{\epsilon_q ^0}{R} - g_{\omega}^q \omega_0 ( {\bf r} ) 
         - (m_q -g_\sigma^q \sigma ( {\bf r} ))
\label{eq:e+-}
\end{eqnarray}
with $\sigma( {\bf r} )$ and $\omega_0 ( {\bf r} )$ being the $\sigma$- and
the time component of the $\omega$-fields.
Since we are dealing with the ground state, Eqs. (\ref{eq:e+-})
are the equations in the radial coordinate $r$ only.
The linear boundary condition from  
Eq. (\ref{eq:psi}), when rewritten by using Eq. (\ref{eq:psi-tr}), reads
\begin{equation}
f \left( \frac{x_q r}{R} \right) = -g \left( \frac{x_q r}{R} \right)
\end{equation}
at the bag surface, $ r=R$, and determines the eigenvalue $x_q$ of the quarks.
$\epsilon_q ^0$ is then given by $\sqrt{ x_q ^2 + (Rm_q )^2 } = x_q$ for
$m_q = 0$.

In the static spherical approximation
Eqs.  (\ref{eq:sigma}) and (\ref{eq:omega}) are reduced to 
\begin{equation}
(\nabla^2 - m_{\sigma}^2 ) \sigma ( {\bf r} )= 
  - g_{\sigma}^q (3 \rho_s )  \theta (R -r)
\label{eq:sigma2}
\end{equation}
and
\begin{equation}
(\nabla^2 - m_{\omega}^2 ) \omega_0 ( {\bf r} ) = 
 - g_{\omega}^q (3 \rho_B ) \theta (R -r),
\label{eq:omega2}
\end{equation}
respectively,
where $\rho_s ( = \bar{\psi}\psi)$ and $\rho_B ( = \bar{\psi} \gamma_0 \psi)$ 
are scalar and baryon densities, respectively, and 
3 is multiplied by them to account for the sum over 3 quarks. 
Equations (\ref{eq:psi1}), (\ref{eq:psi2}), (\ref{eq:sigma2}), and
(\ref{eq:omega2}) constitute a set of
coupled equations for $\psi$, $\sigma$, and
$\omega_0$, which need to be solved self-consistently.

By solving these equations  we can obtain the eigenvalue of the
quarks and the energy ($E_N$) of the nucleon bag. 
$E_{N}$ can be computed by using
\begin{equation}
E_{N} = \int d^3 r \;T_{00}
\end{equation}
with                       
\begin{equation}
T_{00}=\left( \frac{ {\cal E}_q }{R} \bar{\psi} \gamma^0 \psi 
      + B \right) \theta (R -r)
-\frac{1}{2} \left( (\nabla \omega_0 )^2 
+ (m_\omega \omega_0 )^2 \right) 
+\frac{1}{2} \left( (\nabla \sigma )^2 + (m_\sigma \sigma)^2 \right),
\end{equation}
where ${\cal E}_q $ is given by
\begin{equation}
\frac{ {\cal E}_q }{R}  
        = 3 \frac{\epsilon_q ^0 }{R} - \frac{Z}{R} 
        = 3 \frac{x_q        }{R} - \frac{Z}{R} 
\end{equation}
with the sum over 3 quarks taken into account. Correcting for spurious
center-of-mass motion in the bag, the mass of the nucleon bag at rest is
taken to be \cite{Fleck,ST1}
\begin{equation}
M_{N} = \sqrt{E_{N}^2 - \langle p^2_{c.m.} \rangle },
\label{eq:mass}
\end{equation}
where $\langle p^2_{c.m.} \rangle = \sum_{k=1}^3 \langle p_k^2
\rangle = 3 (x_q / R)^2 $. 
By minimizing $M_{N}$ with respect to the bag radius $R$,
we can get the nucleon mass and the bag radius.

\section{Quark-meson coupling model for nuclear matter}

Now, let us consider the QMC model for nuclear matter.
Detailed formulations and justification of the model
can be found in Refs. \cite{Guichon,Fleck,ST1,ST2},
and here we only briefly sketch the model for later discussions.
In addition to the static spherical approximation 
assumed in Section 2 we now further use the mean-field approximation.
From Eq. (\ref{eq:psi}) the quark field $\psi$ inside the bag
in the mean-field approximation may be written as
\begin{equation}
\left[ i \gamma^\mu \partial_\mu - g_\omega^q \bar{\omega} \gamma ^0 
- (m_q - g_\sigma^q \bar{\sigma} ) \right]
\psi  = 0,
\label{eq:psi3}
\end{equation}
where $\bar{\sigma}$ and $\bar{\omega}$ are the mean-fields for
$\sigma$- and $\omega$-mesons, respectively.
Eq. (\ref{eq:psi3}) has
the following simple solution,
\begin{equation}
\psi(t,{\bf r}) \; = \; {\cal N} e^{-i\epsilon_q t/R} 
\left( 
\begin{array}{c}
             j_0 (xr/R) \\
   i \beta_q {\bf \vec{\sigma} } \cdot {\bf \hat{r}} j_1 (xr/R) \\
\end{array}  
\right) \frac{\chi_q }{\sqrt{4\pi}},
\label{eq:psi-tr2}
\end{equation}
where $\epsilon_q = \Omega_q + g_{\omega}^q \bar{\omega}R$,
\begin{equation}
\beta_q = \sqrt{ \frac{\Omega_q - R\; m_q^* }{\Omega_q + R\; m_q ^*} }, \;\; 
{\cal N}^{-2} = 2 R^3 j^2 _0 (x) [ \Omega_q (\Omega_q -1) + R\;m_q ^* /2 ] /x^2
\end{equation}
with $\Omega_q = \sqrt{ x^2 + (R\;m_q ^* )^2 }$ and
$m^* _q = m_q - g_{\sigma}^q \bar{\sigma}$.
The boundary condition
\begin{equation}
j_0 (xr/R) = \beta_q j_1 (xr/R)
\label{eq:j0j1}
\end{equation}
at $r=R$ gives us the $x$ value. Then the energy of the bag can 
be expressed as
\begin{equation}
E_{bag} = 3 \frac{\Omega_q }{R} - \frac{Z}{R} + \frac{4}{3}\pi R^3 B,
\end{equation}
and the effective mass of the bag in nuclear matter is taken as 
\begin{equation}
M_{bag}^* = \sqrt{E^2 _{bag} - \langle p^2 _{c.m.} \rangle  }.
\end{equation}
Using the minimum condition
\begin{equation}
\frac{ \partial M_{bag}^* }{ \partial R} = 0,
\end{equation}
we can determine the effective mass.

The total energy per nucleon at nuclear matter density, $\rho_N$,
including the Fermi motion of the nucleons, can be written as
\cite{Fleck,ST1}
\begin{equation}
E_{tot} = \frac{ \gamma }{ (2\pi)^3 \rho_N }
          \int ^{k_F } d^3 k \sqrt{ M_{bag} ^{*2} + {\bf k}^2 } 
          + \frac{ (3g_{\omega}^q)^2 }{2 m_{\omega}^2 } \rho_N
          + \frac{ m_{\sigma}^2 }{ 2 \rho_N} \bar{\sigma} ^2,
\end{equation}
where the spin-isospin degeneracy $\gamma$ is 4 for symmetric
nuclear matter.

The mean field $\bar{\omega}$ created by the nucleons 
is given by \cite{SW,Guichon,Fleck,ST1}
\begin{equation}
\bar{\omega} = \frac{ 3 g_{\omega}^q \rho_N } { m_\omega ^2 }
\end{equation}
and the scalar mean field $\bar{\sigma}$ 
is determined by the thermodynamic condition
\begin{equation}
\left( \frac{\partial E_{tot} } {\partial \bar{\sigma} } \right)_{R,\; \rho_N}
= 0,
\end{equation}
which yields the self-consistency equation 
\begin{equation}
\bar{\sigma } = \frac{3g_\sigma ^q }{ m_\sigma ^2 } C(\bar{\sigma} )
\frac{\gamma}{(2\pi)^3 } \int^{k_F } d^3 k 
\frac{M_{bag}^* } {\sqrt{M_{bag}^{*2} + {\bf k}^2 }  }
\label{eq:sigmabar}
\end{equation}
with
\begin{equation}
3g_{\sigma}^q C(\bar{\sigma} ) = 3g_{\sigma}^q \frac{E_{bag} }{M_{bag}^*}
\left[ \left( 1- \frac{\Omega_q}{ E_{bag}R } \right) S(\bar{\sigma})
+\frac{m_q ^*}{E_{bag} } \right]
\end{equation} 
and
\begin{equation}
S(\bar{\sigma} ) = \frac{ \Omega_q /2 + R\; m_q ^* (\Omega_q -1 ) }
                  { \Omega_q (\Omega_q -1) + R\; m_q ^* /2 }.
\end{equation}
Equation (\ref{eq:sigmabar}) is also to be solved consistently 
with Eq. (\ref{eq:j0j1}).

\section{Results and Discussions}
We first present the results for a single nucleon.
In presenting the results we take the cases in which the bag radius
is 1 fm.
In Fig. 1 the nucleon mass $M_N$
is  plotted as a function of the bag radius $R$.
The solid  curve represents the free MIT bag mass ($M_N ^{free}$)
without a coupling of the quarks with the mesons; $g_{\sigma}^q =
g_{\omega}^q = 0$. In the calculations
$B^{1/4}$ and $Z$ are taken to be 136.3 MeV
and 1.153 \cite{ST1}, respectively, which give us the minimum 
of $M_N ^{free}$ as 939 MeV at 1.0 fm. 
If we include the interaction of quarks with meson fields using the
coupling constants $g_{\sigma}^q = 5.605$ and $g_{\omega}^q = 1.152$
chosen by Saito and Thomas\cite{ST1}, for instance, 
a solution to the field equations 
(\ref{eq:psi1}), (\ref{eq:psi2}), (\ref{eq:sigma2}), and
(\ref{eq:omega2})  does not exist.   
In Eq. (\ref{eq:psi2}) both $dg(r)/dr$ and $f(r)$ are
negative (or zero) for all radii, and thus $E^+ (r)$ needs to be positive.
However, due to the large attraction coming from the large $g_{\sigma}^q$
value in relative to the small $g_{\omega}^q$ 
the eigenvalue $\epsilon_q ^0$ in Eq. (\ref{eq:e+-}) 
becomes relatively small, which results in $E^+ (r)$ 
being negative for this set of parameters;
hence the non-existence of the solution.
Therefore we show here results for other sets of parameters.

First, we include only the
coupling of the quarks with the $\sigma$-meson; $g_{\omega}^q =0$. 
The bag mass
decreases due to the attractive $\sigma$-coupling as shown in Fig. 1 by the 
thin dashed and the thin dash-dotted curves, respectively, for
$g^q_{\sigma} = $ 2 and 4. 
$M_N$ is reduced to 930 and 903 MeV, and the bag radius also  
slightly decreases to 0.997 and 0.979 fm, respectively, for
$g_{\sigma}^q = 2$ and 4.
On the other hand, when the quarks are allowed to couple only with
the $\omega$-field, {\it i.e.,} $g_{\sigma}^q =0$, $M_N$ increases as
plotted  by the thick dashed and the thick dash-dotted curves, 
respectively, for $g_{\omega}^q = 2$ and $4$.
$M_N$ is pushed up to 954 and 994 MeV 
due to the repulsive coupling,
and the bag radius becomes 1.00 and 1.03 fm, respectively.

If the quarks are allowed to couple with
both the $\sigma$- and the $\omega$-mesons, cancellation between
the attraction and the repulsion takes place.
In Fig. 2 we show $M_N$ calculated with
$g_{\sigma}^q = g_{\omega}^q =2$ 
by the dashed curve in comparison with $M_N ^{free}$
($g_{\sigma}^q = g_{\omega}^q =0$) plotted by the solid curve.
$M_N$ for  $g_{\sigma}^q = g_{\omega}^q =2$ has a minimum at
946 MeV and $R=1.01$ fm. The mass difference ($946-939=7$ MeV)
is just about what is left after the cancellation
between the thin dashed curve ($g_{\sigma}^q =2$, $g_{\omega}^q =0$) 
and the thick dashed curve ($g_{\sigma}^q =0$, $g_{\omega}^q =2$) in Fig. 1.
$M_N$ calculated with $g_{\sigma}^q = g_{\omega}^q = 4$ is also plotted
in Fig. 2 by the dash-dotted curve. The minimum of $M_N$ is
964 MeV at $R=1.02$ fm. Again, the mass shift ($964-939 = 15$ MeV)
is just about the repulsion left after the cancellation between
the two dash-dotted curves in Fig. 1. 
It is worth noting here that when we use the same value for
$g_{\sigma}^q$ and $g_{\omega}^q$ the net result is a repulsion.
This may be understood by looking at the meson fields. 
In Fig. 3 $\sigma ( {\bf r} )$ and 
$\omega_0 ( {\bf r} )$ are plotted as a function
of radius for a fixed bag radius $R = 1$ fm
when $g_{\sigma}^q = g_{\omega}^q = 2$.
The figure shows that the shapes of 
$\sigma ( {\bf r} )$ and $\omega_0 ( {\bf r} )$
are largely dictated by those of the source densities $\rho_s $ and $\rho_B$,
respectively.
Just as $\rho_B $ is bigger than $\rho_s$ at larger radii,
$\omega_0 ( {\bf r} )$ is greater than $\sigma ( {\bf r} )$ at larger
radii, which is the significant radial region.
Thus, when the two coupling constants are equal,  
a net effect turns out to be repulsive. 

In the nuclear matter at normal nuclear densities
the internucleon distance is about 2 fm. 
In the mean-field approximation meson fields are assumed to be constant 
in the matter.
On the other hand, Fig. 3 (a) shows that 
the range of the meson fields is roughly around 
1.5 fm for both mesons. The fields become almost negligible 
at $r \approx 2$ fm. However, since the nucleons in nuclear matter
are in the Fermi motion, 
the meson fields with the range of about 1.5 fm may be
approximated to mean-fields in the matter.
Therefore, the present study of a single nucleon in the QMC model
seems compatible with the QMC model for the nuclear matter, 
which will be discussed shortly.

We then looked for parameters that could give us the correct nucleon
mass for $R=1$ fm. First we fixed $B^{1/4}$ and $Z$ as 136.3 MeV and 
1.153 \cite{ST1}, respectively,
and searched for $g_{\sigma}^q$ and $g_{\omega}^q$.
It was not possible to find $g_{\sigma}^q$ and 
$g_{\omega}^q$ that would yield the correct nucleon mass
with the given values of $B^{1/4}$ and $Z$.
When we let $Z$ also vary, keeping $B^{1/4}$ fixed as 136.3 MeV, 
it was still
impossible to find parameters producing the correct nucleon mass.
On the other hand, when we let $B^{1/4}$ vary, keeping $Z$ as 1.153,
we could obtain the correct mass with the parameter set 2 listed in Table 1.
There are four parameters to be fixed, whereas there are only two
quantities to fit; the nucleon mass and the bag radius.
Thus the parameters are not unique. 
We list in Table 1 other sets of parameters that can produce the
nucleon mass of 939 MeV for $R =1$ fm.
$M_N$'s calculated with the parameter sets 2 - 7 are all 
indistinguishable from the solid curve in Fig. 1 or Fig. 2, so
they are not plotted here.
One can see from Table 1 that $g_{\sigma} ^q$ is 
not much larger than $g_{\omega}^q$.
$g_{\sigma}^q$ is often 
only slightly bigger than $g_{\omega}^q$, 
as the relativistic Dirac phenomenolgy has shown. 
In fact, this is expected from Fig. 2, which shows that with the equal
values of $g_{\sigma}^q$ and $g_{\omega}^q$ there is only 
a small repulsion. To restore the correct mass,
$g_{\sigma}^q$ would have to be a little larger than $g_{\omega}^q$.
This is in contrast to the values of $g_{\sigma} ^q$ 
and $g_{\omega}^q$ previously obtained by QMC model calculations for
nuclear matter. Too large $g_{\sigma} ^q$ values compared to 
small $g_{\omega} ^q$ were one of the concerns in the previous results of
the QMC model calculations. 

Similar studies were made at other bag radii ranging from 0.6 to 1.0 fm,
and similar results were obtained. 
Our results show that the change in the nucleon 
mass due to the  coupling of the quarks with the mesons is not
negligible. 
Such self-energy effects 
on the nucleon mass need to be taken into consideration
in choosing the parameters for 
the calculation of nuclear matter properties.
Thus, we performed the nuclear matter calculations as sketched in
Section 3 along with the calculation for a nucleon. 
As mentioned earlier, when we use 
$g_{\sigma} ^q$ and $g_{\omega}^q$
taken from Ref. \cite{ST1}, a solution for a nucleon problem does not exist.
On the other hand, if we use the parameter sets 2 - 7 in Table 1,
the calculated binding energies are not correct. 
Even saturation does not occur for these parameter sets.  
Therefore, we attempted to look for a new set of parameters that
could reproduce simultaneously the nucleon mass 
and the nuclear binding energy.
We found that it was not possible to get such parameters, at least, 
within the present scheme.
Therefore, this additional requirement of reproducing a single nucleon mass
can be used as a guide in 
refining the QMC model for nuclear matter.

\section{Summary}
Recently the QMC model has been frequently used to describe 
the properties of nuclear matter and finite nuclei.
However, in the previous calculations 
the change in the mass of the
bag due to the self-energy has been ignored.
Thus we have applied the QMC model to a single nucleon.
Our calculations suggest that this change in the nucleon mass is not
negligible. Therefore, we performed the QMC model calculations 
not only for a nucleon but also for 
nuclear matter and looked for parameters that would
give us both the nucleon mass and 
the nuclear binding energy correctly.
However, we only found that it was not possible to obtain such parameters
within the present framework of the model.
Taking into account this change in the nucleon mass may provide us
with an additional constraint on the model.

\section{Acknowledgements}

This work was supported by the Natural Sciences and Engineering Research
Council of Canada.
SWH acknowledges kind hospitality at TRIUMF. He also thanks Prof. T. Udagawa
for helpful discussions and was partially supported 
by Sungkyun Faculty Research Fund and BK21 Physics Research Division.

\pagebreak

\renewcommand
\arraystretch{1.8}
\begin{tabular}{c|cccccc}    \hline\hline

~set~ & ~$B^{1/4}$(MeV)~~&~~$Z$~~~&~~$g_{\sigma}^q$ ~&~~$g_{\omega}^q$ & 
$M_N$ (MeV)      \\ \hline
  1 & 136.3  & 1.153 & 5.605  & 1.152 & No solution.  \\ \hline
  2 & 136.5  & 1.153 & 3.396  & 2.490 & 939          \\ \hline
  3 & 136.5  & 1.169 & 3.300  & 2.616 & 939          \\ \hline
  4 & 136.5  & 1.183 & 3.199  & 2.723 & 939          \\ \hline
  5 & 136.5  & 1.199 & 2.941  & 2.760 & 939          \\ \hline
  6 & 136.5  & 1.170 & 1.965  & 1.752 & 939          \\ \hline
  7 & 136.4  & 1.238 & 3.840  & 3.746 & 939          \\ \hline
\end{tabular}

\vspace{1cm}

\begin{table}
\baselineskip=24ptplus.5ptminus.2pt

\caption{The parameter set 1 is taken from Ref. [6].
The paremeter sets 2 - 7 are the parameters that can produce
$M_N = 939$ MeV at the bag radius $R=1$ fm.
These parameters are not unique, so here we list some of them.}


\end{table}
\pagebreak

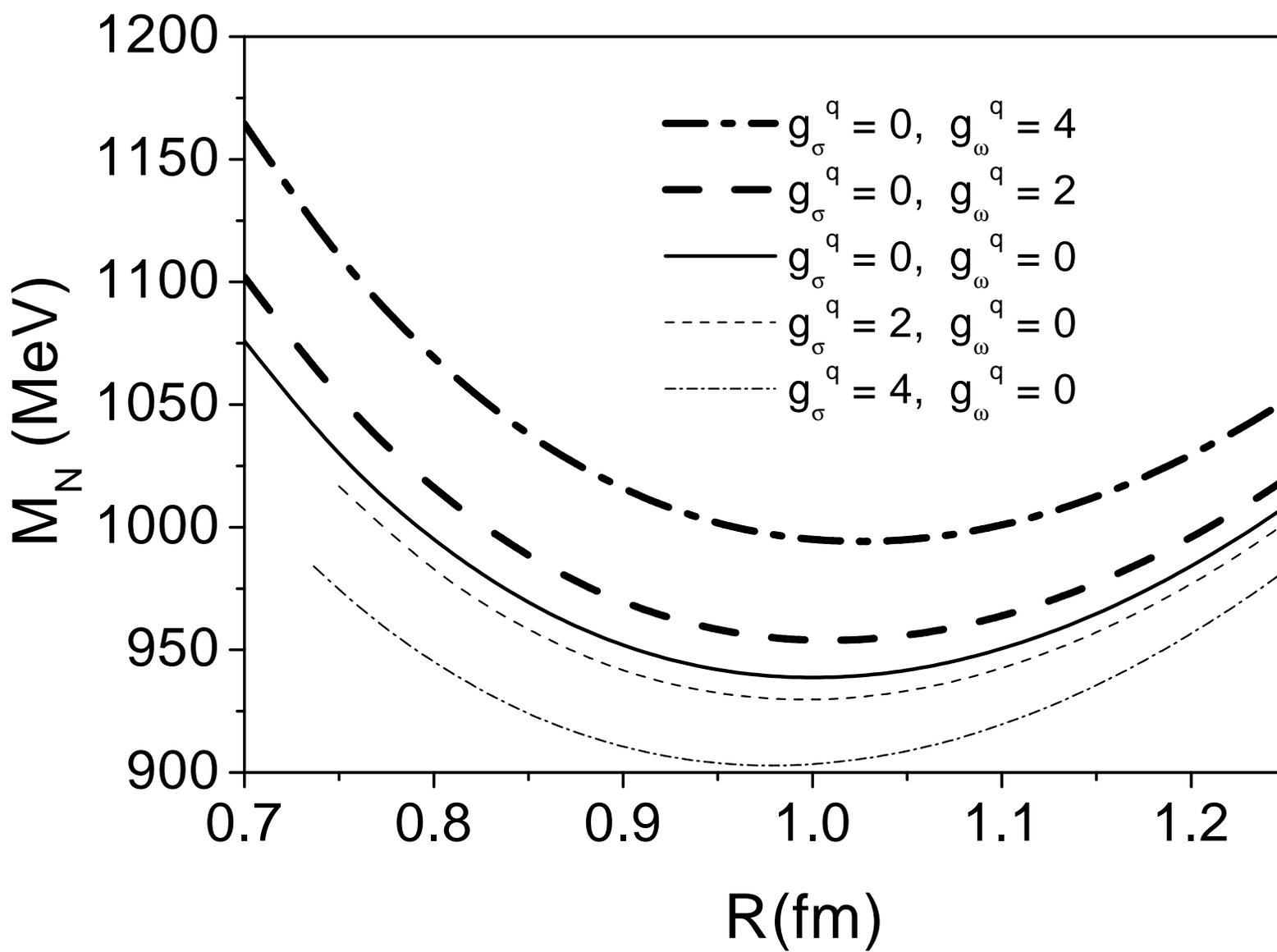
\begin{figure}
\caption{The nucleon mass ($M_N$) calculated by the QMC model for a nucleon
is plotted as a function of the bag radius $R$
for different coupling constants $g_{\sigma}^q$ and $g_{\omega}^q$.
$B^{1/4} = 136.3$ MeV and $Z = 1.153$ are used.} 
\end{figure}

\begin{figure}
\caption{The nucleon mass ($M_N$) calculated by the QMC model for a nucleon 
is plotted as a function of the bag radius $R$
for different coupling constants $g_{\sigma}^q$ and $g_{\omega}^q$. 
$B^{1/4} = 136.3$ MeV and $Z = 1.153$ are used.} 
\end{figure}

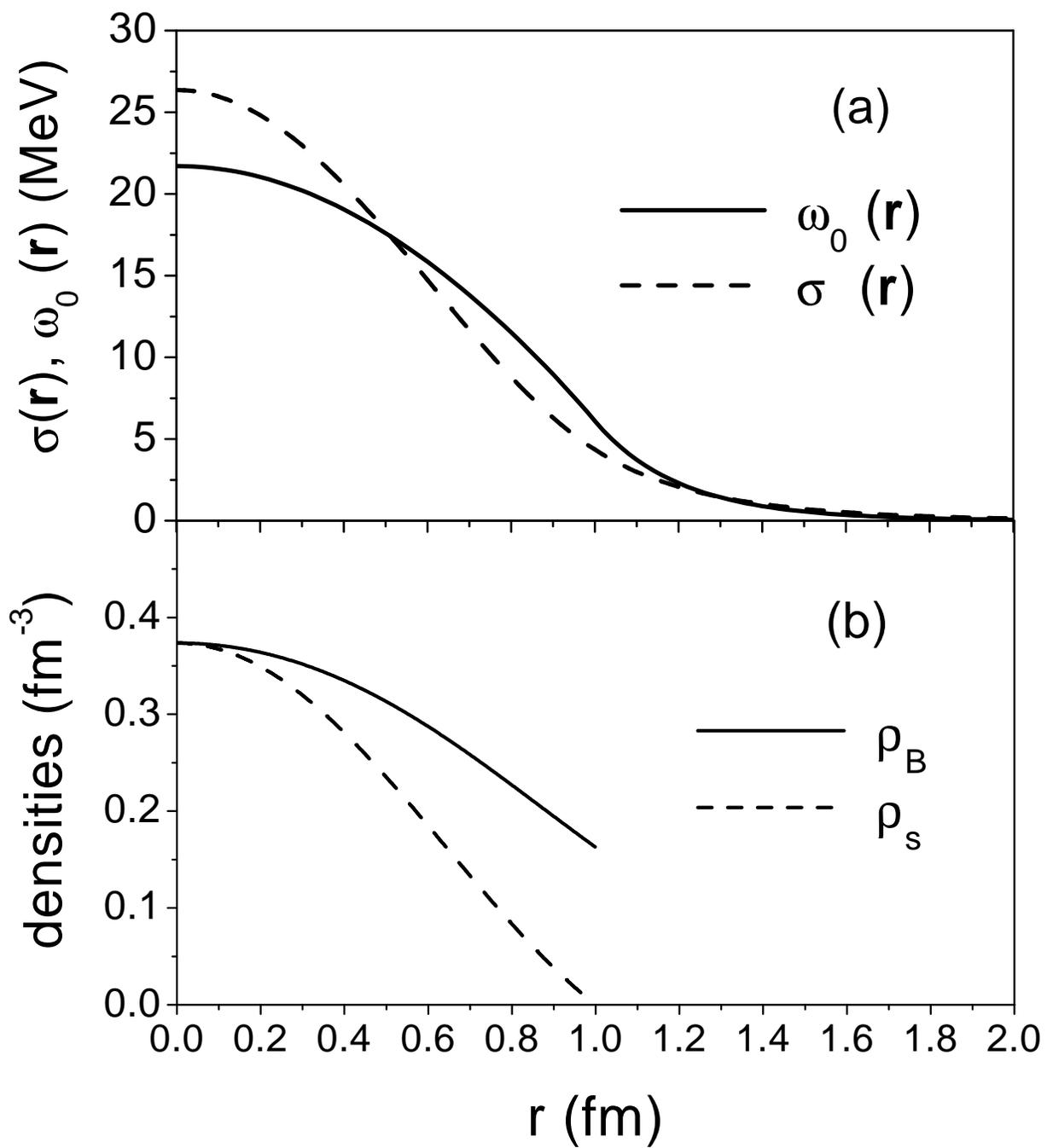
\begin{figure}
\caption{(a) The $\sigma$- and the $\omega$-meson fields 
calculated by the QMC model for a nucleon 
are plotted as a function of radius $r$ for bag radius $R =1$ fm
with $g_{\sigma}^q = g_{\omega}^q = 2$, $B^{1/4}=136.3$ MeV and 
$Z = 1.153$. (b) The scalar and the baryon densities are
plotted by the dashed and the solid curves, respectively,
with the same parameters.}
\end{figure}

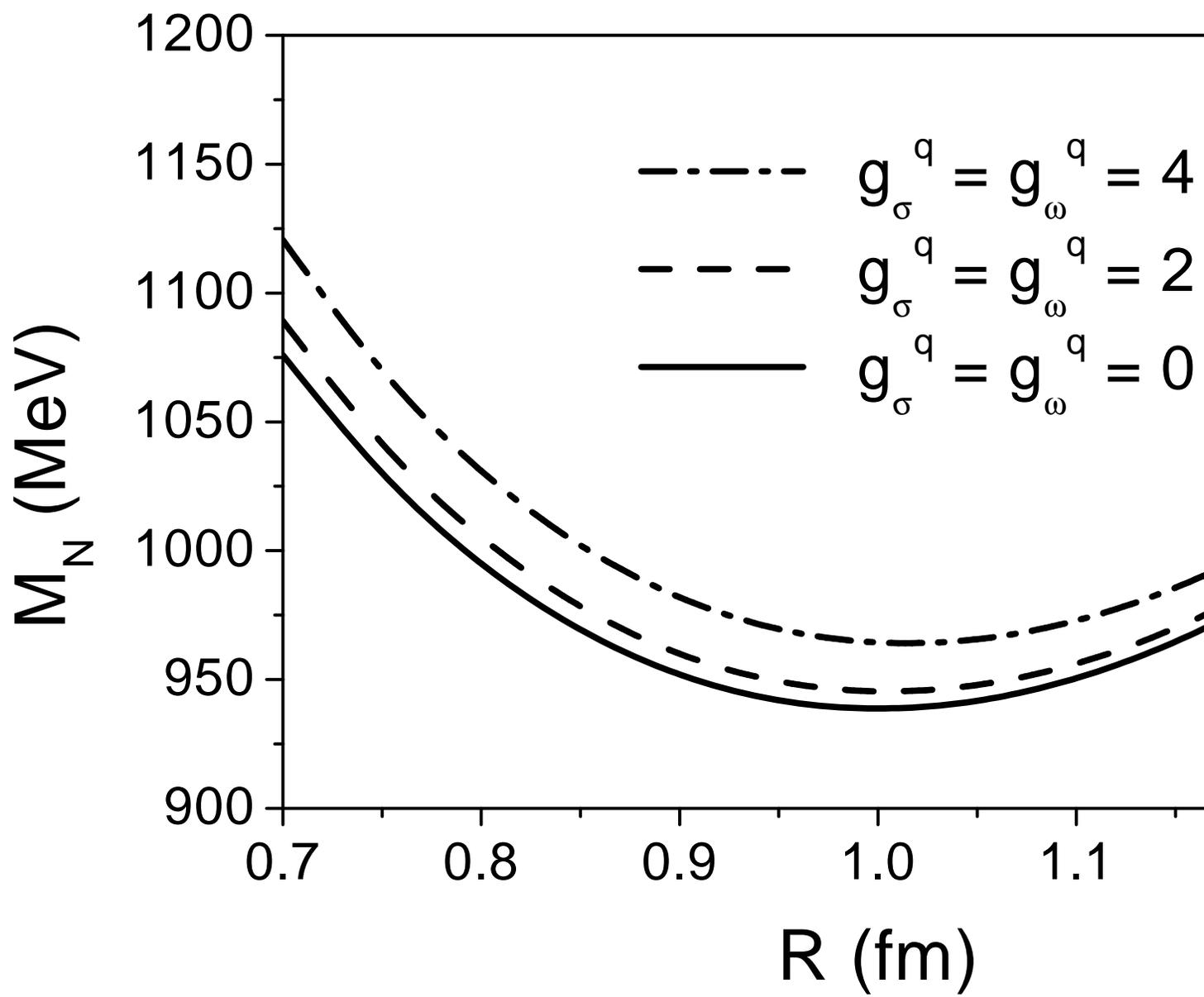
\end{document}